\documentstyle[editedvolume]{crckapb} 
\input psfig.sty

\begin{opening}
\title{ X-ray variability and spectral scaling: 
\protect\\ a measure of BLR sizes in AGN}

\author{A. Wandel}
\institute{Racah Institute, The Hebrew University, Jerusalem, Israel}
\author {Th. Boller}
\institute{Max Planck Inst. f\"ur Extraterr. Physik, Garching, Germany}

\end{opening}

\runningtitle{ BLR sizes in AGN}

\begin{document}


\begin{abstract}
We developed a new method of determination of the size of the broad
emission-line region (BLR) in active galactic nuclei. This method
relates the radius of the broad-line region of AGN to the soft X-ray
luminosity and spectral index.
Comparing the BLR distances calculated from our
model to the BLR distances determined by reverberation mapping shows that
our scaling law agrees with the $R\sim L^{1/2}$ empirical relation.
Here we investigate  a complimentary method of estimating the BLR distance - 
based on the Keplerian broadening of the emission lines
and the central mass estimated from X-ray variability.
\end{abstract}

\section{Introduction}

Recent results from reverberation-mapping of the broad emission-line regions 
(BLR) in AGN indicate that the BLR distance from the central radiation source
roughly scales as $r\propto L^{1/2}$ (Peterson 1995).
Recently we have elaborated a different method for estimating the BLR distance,
using the emission-line photoionization model.
In order to explain the anticorrelation between the H$\beta$ line width 
and the soft X-ray spectral slope in  Narrow-Line Seyfert 1 galaxies (NLS1) 
 (Boller, Brandt and Fink 1996; Wang, Brinkmann and Bergeron 1996),
 Wandel and Boller (1997) showed that a steeper spectrum
would produce a BLR at a relatively
larger distance from the central source, so the Keplerian velocity is lower.
This modified BLR distance calculated from the photoionization relation
agrees well with the reverberation distance (Wandel 1996; 1997).
In this work we connect this model to the r(BLR)~$L^{1/2}$ relation by
independently  determining the central mass from the X-ray variability.

\section{The Line-width spectral slope correlation}

 If the emission lines are Doppler-broadened by  Keplerian motion in the
  gravitational potential of the central mass, the full width at half
maximum  is given by:
$$ FWHM \approx \left(\frac{GM}{R}\right)^{\frac {1}{2}}    \eqno(1)  $$
where M is the mass of the central black hole
and R the radius of the emission line region.
The physical conditions in the ionized gas emitting the broad lines are
characterized by the ionization parameter U, the ratio of ionizing photons to
electrons 
$U =\int_{E_0}^\infty f(E) \frac{dE}{E} /{4\pi R^2} c n_e$
where $f(E)$ is the luminosity of the central source, per unit
energy, $n_e$ is the electron density.
The radius of the BLR may then be written as
$$
R= \left( \frac{L_{ion}}{ 4\pi c\bar E_{ion} U n_e}\right ) ^{1/2}
\eqno (2)
$$
where
$L_{ion}$ is the ionizing luminosity,
and $\bar E_{ion}$
is the mean energy of the ionizing photons.
%
Typical values 
in AGN clouds give  $Un \sim 10^9-10^{11} cm^{-3}$
(cf. Rees, Netzer \& Ferland 1989).
For $nU = 10^{10}$ eq. (2) gives
$$ 
R \simeq 0.037 \bar E_{ion}^{-1/2} \left ({ L_{ion}\over 10^{45}erg/s}
\right ) ^{1/2}~~ pc \eqno (3) 
$$
where $\bar E_{ion}$ is in Rydbergs.
Assuming the ionizing spectrum has a power-law shape with the spectral
index of the soft X-ray band (found e.g. from the ROSAT data), 
 $f(E) \propto E^{-\alpha}$ , gives
the observed anti-correlation between the line width and the
spectral index (Wandel and Boller 1997).

\section{The radius-luminosity relation}

Wandel (1996; 1997) has shown that the BLR radius calculated from eq. (2) above
does indeed agree well with the radius estimates available from
reverberation calculations for about a dozen objects (see Kaspi et.al. 1996 and
these proceedings).
We present  an independent derivation of the 
 radius-luminosity relation, determining the mass from
X-ray variability. 

\begin{figure}
        \psfig{figure=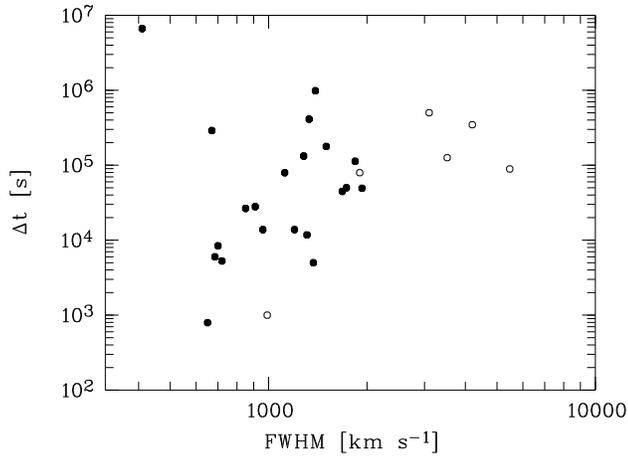,height=10cm}
      \caption{X-ray doubling times
               versus FWHM of H$\beta$ for Seyfert 1 galaxies (open circles) 
	and NLS1 (filled circles).}
\end{figure}
An upper limit for the black hole mass is given by (e.g. Wandel \&
Mushotzky 1986)
$$M(\Delta t) < (c^3 / 10 G) \Delta t \simeq 10^4 \Delta t\ M_{\odot}, \eqno
(4)$$
where $\Delta t$ is in seconds.
This relation assumes that the bulk of the X-ray continuum is emitted within 5
Schwarzschild radii.
Using the doubling time as the characteristic time for variability we
determine upper limits for the black hole mass for those objects
 whose continuum luminosities have been
observed to vary significantly.
Fig. 1 gives the doubling time (we have extrapolated amplitude
variations linearly to a factor of 2 to determine $\Delta t$)  versus the observed
FWHM. Although there is some scatter, there is a strong indication for an
increase of the doubling time with observed FWHM.
This may indicate a relation between the central mass and the line width.
We may use eq. (1) to establish an estimate for the BLR
radius calculated from the observed line width and X-ray variability,
independent of the luminosity, which can test the R-L$^{1/2}$ relation:
$$R\approx GM/v^2 < (1.3 10^{17} cm) \left({\Delta t}\over{10^4
sec}\right ) \eqno (5)$$
Note that this is an upper limit, as is eq. (4).
The diamonds in figure 2 show the BLR radius calculated from eq. (5)
vs. the X-ray luminosity for
the objects in our sample with significant X-ray variability
and a few objects from Wandel and Mushotzky (1986).
We see that there is no correlation between R and L.
On the other hand, as expected, the BLR radius calculated from the 
photoionization model (eq. 3)  shows a clear
$R_{BLR}\sim L_x^{1/2} $ correlation (circles in figure 2), 
which is confirmed independently by reverberatiom mapping.
Looking at figure 2 we may conclude that the X-ray variability method
systematically overestimates the central mass for the low luminosity
NLS1 galaxies.

\begin{figure}
	\psfig{figure=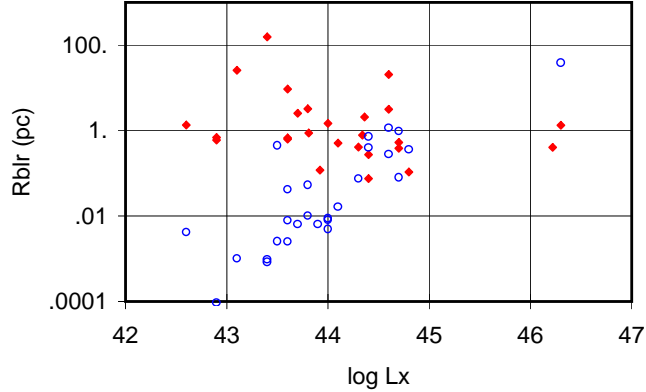,height=9.cm}
      \caption{BLR radii calculated from the X-ray doubling times (diamonds)
               and from the  H$\beta$ emission lines (circles ) vs. X-ray 
		luminosity.}
\end{figure}

{}


\begin{thebibliography}{}
 

\bibitem[]{} Boller, Th., Brandt, W.N., Fink, H. 1996,  A\&A, 305, 53

\bibitem[]{}Kaspi, Sh. et.al. 1996, ApJ , 471, L75

\bibitem[]{}Peterson, B.M. in "Reverberation Mapping of AGN" eds. P.M. Gondhalekar, K.Horne, B.M.Peterson, SFASP 1995.

\bibitem[]{}Rees, M., Netzer H., Ferland, G.J. 1989. ApJ , 347, 640.

 
\bibitem[]{}Wandel, A. and Mushotzky, R.F. 1986, ApJ  , 306, L61.


\bibitem[]{}Wandel, A. 1996, in "X-ray Imaging and Spectroscopy of Cosmic
Hot Plasmas", ed. F.Makino.

\bibitem[]{}Wandel, A. 1997, ApJ Letters, in press.
\bibitem[]{}Wandel, A. and Boller, Th.. 1997, A\&A , in press.
\end{thebibliography}
\end{document}